\documentclass[english]{lni}
\newcommand{\PaperAcronym}{GLLM\xspace}
\newcommand{\quotes}[1]{``#1''}
\usepackage{algorithm}
\usepackage{algpseudocode}
\usepackage{amsmath}
\usepackage{subfig}
\begin{document}
\title{\PaperAcronym{}: Self-Corrective G-Code Generation using Large Language Models with User Feedback}
 \author[1]{Mohamed Abdelaal}{Mohamed.Abdelaal@softwareag.com}{}
 \author[2]{Samuel Lokadjaja}{Sammyloka@yahoo.com}{}
 \author[2]{Gilbert Engert}{G.Engert@ptw.tu-Darmstadt.de}{}
 \affil[1]{Software AG\\Uhlandstrasse 12\\64297 Darmstadt\\Germany}
 \affil[2]{TU Darmstadt, Darmstadt, Germany}
\maketitle

\begin{abstract}
This paper introduces \PaperAcronym, an innovative tool that leverages Large Language Models (LLMs) to automatically generate G-code from natural language instructions for Computer Numerical Control (CNC) machining. \PaperAcronym addresses the challenges of manual G-code writing by bridging the gap between human-readable task descriptions and machine-executable code. The system incorporates a fine-tuned StarCoder-3B model, enhanced with domain-specific training data and a Retrieval-Augmented Generation (RAG) mechanism. GLLM employs advanced prompting strategies and a novel self-corrective code generation approach to ensure both syntactic and semantic correctness of the generated G-code. The architecture includes robust validation mechanisms, including syntax checks, G-code-specific verifications, and functional correctness evaluations using Hausdorff distance. By combining these techniques, \PaperAcronym aims to democratize CNC programming, making it more accessible to users without extensive programming experience while maintaining high accuracy and reliability in G-code generation.

\end{abstract}
\section{Introduction}\label{sec:intro}

G-code, a domain-specific programming language, forms the backbone of computer numerical control (CNC) in manufacturing~\cite{autodesk2014fundamentals}. This language dictates the precise movements of machine tools, guiding processes like cutting, drilling, and milling that transform raw materials into finished products. Unlike high-level programming languages, G-code operates at a low level, akin to assembly languages, and lacks an inherent semantic understanding of the geometry being manufactured. Each command, typically initiated with a letter like \quotes{G} or \quotes{M} followed by specific parameters, modifies the state of the machine tool, culminating in a sequence of instructions that orchestrate the entire fabrication process.

 Manually writing G-codes for CNC machines is a challenging and error-prone process. It requires precise coordination of multiple machine axes and careful planning of tool paths to avoid collisions and optimize machining efficiency. Programmers must manually calculate coordinates, speeds, and feed rates for each movement, which is time-consuming and susceptible to errors. The process involves extensive testing and debugging, often requiring multiple revisions to achieve the desired results. Human errors, such as typos or incorrect calculations, can lead to significant issues in the final product. Additionally, creating effective G-code requires specific technical knowledge of machining principles and familiarity with the particular CNC machine being used. These factors combine to make manual G-code writing a complex task that demands expertise and attention to detail, highlighting the potential benefits of automated or AI-assisted programming approaches in modern manufacturing environments.

Existing non-AI solutions for automatic G-code generation primarily revolve around Computer-Aided Manufacturing (CAM) software and slicing software for 3D printing. CAM software packages like Fusion 360~\cite{fusion2024}, Mastercam \cite{mastercam2024}, and Siemens ShopMill \cite{siemens2008shopmill} typically offer user interfaces with pre-defined templates for creating various machining operations without extensive programming knowledge. However, these solutions often present challenges, including time-consuming setup processes, steep learning curves for beginners, high costs associated with software licenses, and limited flexibility due to vendor-specific requirements. On the other hand, slicing software for 3D printing, such as PrusaSlicer and Cura~\cite{cura2024}, provides another avenue for G-code generation, but its scope is limited to additive manufacturing processes and lacks functionality for subtractive manufacturing techniques such as milling and drilling. Additionally, most slicing software offers only basic G-code generation capabilities, which may not suffice for complex machining operations. While these non-AI solutions have been instrumental in automating G-code generation to some extent,  they still require significant human intervention and expertise, leaving room for more advanced, AI-driven approaches to address their limitations and enhance overall efficiency in CNC programming.

\begin{figure}
    \centering
    \includegraphics[width=0.9\linewidth]{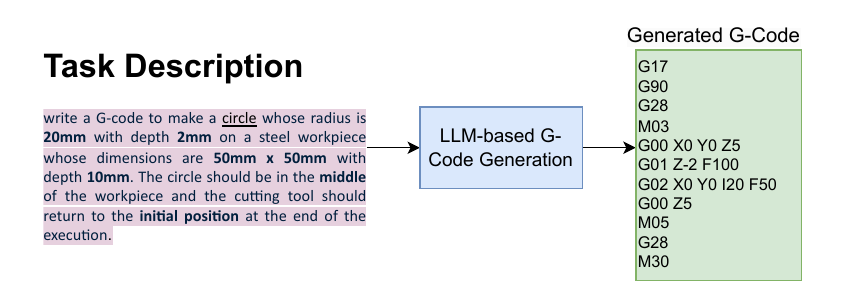}
    \caption{Leveraging LLM to transform natural language instructions into executable G-code, enabling more intuitive programming of CNC machines.}
    \label{fig:idea}
\end{figure}

To overcome these challenges, we introduce \PaperAcronym{}, a novel tool that harnesses the power of Large Language Models (LLMs) to generate G-code from natural language instructions automatically. \PaperAcronym{} bridges the gap between natural language instructions and machine-executable G-code, as depicted in Figure~\ref{fig:idea}, enabling users to express their design intent in plain language and receive accurate, ready-to-use G-code output. By leveraging the advanced natural language understanding capabilities of LLMs, \PaperAcronym{} can interpret complex design descriptions and translate them into precise machining instructions. This approach streamlines the G-code generation process and democratizes CNC machining by making it more accessible to a wider range of users, including those without extensive programming or CAM software experience. Furthermore, \PaperAcronym{} facilitates rapid prototyping and exploration of complex designs, allowing for quick iterations and modifications based on natural language input. As a result, \PaperAcronym{} represents a significant step forward in automating and simplifying the CNC programming workflow, potentially transforming the landscape of digital manufacturing.

To sum up, the paper provides the following contributions: (1) Framework Development: We developed a framework to translate natural language directly into executable G-code. By leveraging open-source large language models (LLMs), we ensure full control over the system and maintain privacy, avoiding the constraints of proprietary models. (2) Enhanced Model Training: Recognizing that LLMs trained on general text data often struggle with the specific syntax and formatting of G-code, we incorporated domain-specific training data. We fine-tuned these models on G-code and manufacturing-related information. Moreover, we implemented a Retrieval-Augmented Generation (RAG) mechanism to improve accuracy. (3) Advanced Prompting Strategies: We designed a novel prompting strategy and input representation to effectively guide LLMs toward accurate G-code generation. (4) Verification of G-code Correctness: To address the challenge of verifying generated G-code accuracy, we developed methods ensuring both syntactic and semantic correctness. This includes a Self-corrective Code Generation mechanism with checks for syntax, continuity, and drilling, alongside a user-in-the-loop approach for validating the expected tool path. To the best of our knowledge, \PaperAcronym{} represents the first work that directly translates natural language to executable G-code.



\section{Architecture \& Overview}\label{sec:intro}

\begin{figure}
    \centering
    \includegraphics[width=1\linewidth]{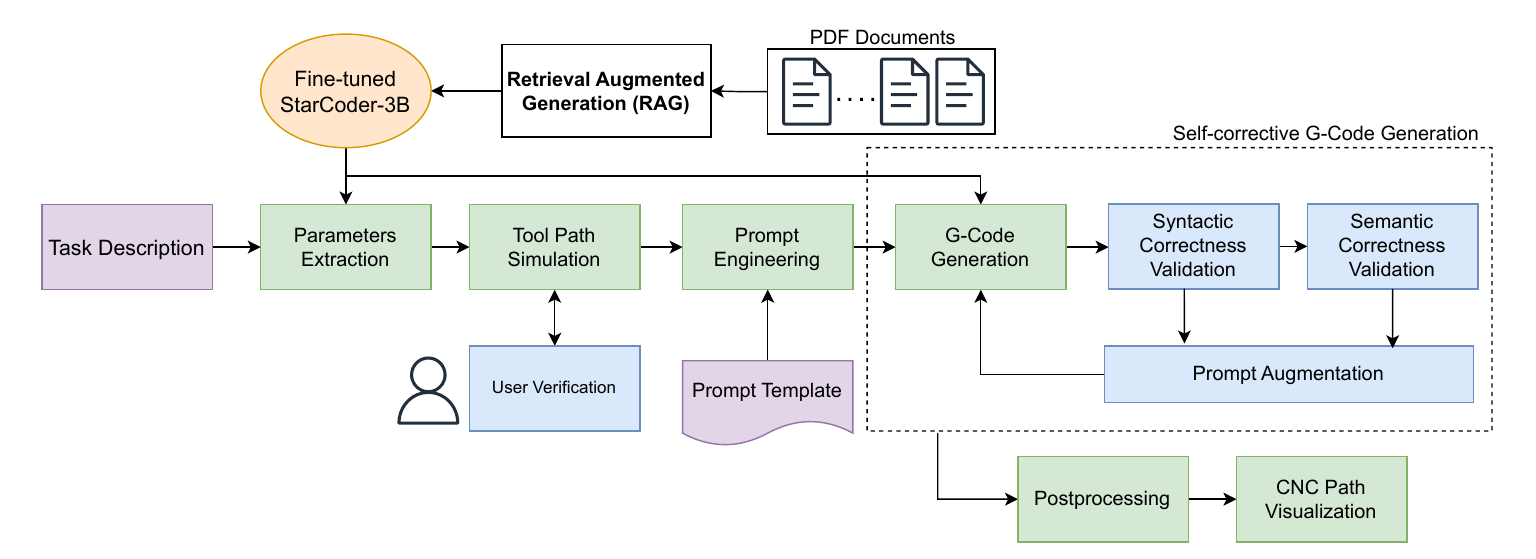}
    \caption{Architecture of \PaperAcronym{}}
    \label{fig:pipeline}
\end{figure}

In this section, we introduce the architecture of \PaperAcronym{}. Figure~\ref{fig:pipeline} depicts the architecture of \PaperAcronym{} designed to generate executable G-code from natural language instructions. The process begins with a user providing a natural language description of the desired machining task. This description is then processed through our fine-tuned StarCoder-3B model, a large language model specifically trained on a G-code dataset extracted from the \quotes{Stack} dataset of public GitHub repositories. To expedite the training process, we employed techniques like parameter efficient fine-tuning (PEFT) and mixed precision training, enabling us to leverage the power of large language models efficiently.

Once the model receives the task description, it moves to the \textit{Parameters Extraction} stage, where crucial information such as workpiece dimensions, cutting depths, and coordinate points are identified. This extracted information forms the basis for both Tool Path Simulation and Prompt Engineering. The Tool Path Simulation module provides an initial visualization of the intended cutting trajectory, allowing for early detection of potential errors or misinterpretations. Concurrently, Prompt Engineering refines the user's input by incorporating a predefined Prompt Template and leveraging a Retrieval Augmented Generation (RAG) mechanism. This mechanism enriches the model's understanding by accessing relevant information from external PDF documents. These documents include comprehensive definitions of various G-code commands and detailed documentation of the target CNC machine, ensuring the generated code aligns with the specific capabilities and constraints of the manufacturing setup.

The refined input, enriched with contextual information from the RAG process, is then passed to the G-Code Generation module. Here, the model utilizes its learned knowledge to produce an initial draft of the G-code program. However, to guarantee the accuracy and reliability of the generated code, it undergoes rigorous validation through a self-corrective G-Code generation mechanism. This mechanism encompasses both Syntactic and Semantic Correctness Validation stages. Syntactic validation checks the generated code for adherence to G-code syntax rules, ensuring each command is correctly structured and free from errors. Semantic validation, on the other hand, focuses on the functional correctness of the generated G-code. To this end, the generated tool path is compared against the user-defined trajectory using similarity measures, specifically the Hausdorff distance. This quantifies the difference between the two paths, highlighting potential discrepancies. The identified errors are then used to augment the initial G-code generation prompt. This feedback loop helps guide the LLM model to avoid previously identified errors, iteratively refining the generated G-code toward a  syntactically and semantically correct solution.

Once the generated G-code successfully passes both validation stages, it proceeds to the Postprocessing step. This step involves two key processes: G-Code Extraction and Parameters Adjustment. G-code extraction filters the output from the model, removing any extraneous text and isolating the essential G-code commands. Parameters Adjustment ensures that user-defined parameters, such as spindle speed and feed rate, are correctly parsed and incorporated into the final G-code program. Finally, the generated and validated G-code undergoes visualization using two methods: a dedicated G-code simulator called CAMotics and a custom Python script designed to plot the cutting path, allowing for final verification and ensuring the code accurately reflects the user's desired machining outcome.

\section{Prompt Engineering}\label{sec:prompt}
%
This section details our systematic approach for transforming raw user prompts into structured input for our G-code generation system. This structured format is crucial for enabling the LLM model to accurately interpret user requirements and generate valid, executable G-code that precisely reflects the desired machining outcome. Specifically, \PaperAcronym{} involves extracting essential parameters from a user's CNC machining task description and managing missing information. The process begins with the user providing a description of their desired machining task. This description is then fed into a fine-tuned StarCoder-3B model that parses the text, identifying key parameters such as material type, operation type (e.g., milling, drilling), desired shape, workpiece dimensions, starting point, home position, cutting tool path, whether to return the tool to the home position after execution, depth of cut, feed rate, and spindle speed.

The extracted parameters are then organized into a structured Python dictionary, facilitating further processing. This dictionary is then compared against a predefined G-code prompt template that outlines all the necessary parameters for G-code generation. This template serves as a checklist, ensuring all required information is present before proceeding. \PaperAcronym{} detects missing parameters by carefully analyzing the populated parameter dictionary. If any parameters are found to be missing, the system prompts the user to provide the undefined values. This interactive feedback loop ensures the system has all the necessary information to generate a complete and executable G-code program. The extracted information is then visualized, typically as a 2D representation of the toolpath, and presented to the user for verification. This step ensures the system correctly interprets the user's intent before proceeding with G-code generation.

To further enhance \PaperAcronym{}'s capabilities, we introduce a novel prompt design that facilitates the automatic determination of the number of distinct shapes within a user's task description. This enables \PaperAcronym{} to handle more complex machining operations involving multiple geometric features. For instance, if a user requests G-code for a rectangular pocket featuring two internal islands (cf. Task \#5 in Section~\ref{sec:evaluation}), \PaperAcronym{} infers the presence of three separate shapes. To address this multi-shape scenario, \PaperAcronym{} employs a \textit{description decomposition} strategy. This strategy deconstructs the original task description into individual subtask descriptions, each corresponding to a single shape. This decomposition allows \PaperAcronym{} to apply its parameter extraction and self-corrective G-code generation mechanisms to each subtask in isolation. This focused approach ensures accurate G-code generation for each individual geometric element. Following the generation of subtask-specific G-code segments, \PaperAcronym{} performs a refinement and integration step. This crucial step combines the individual code segments into a unified and executable G-code program. This final program accurately reflects the complete user-defined task, encompassing all specified shapes and their relationships. This decomposition and integration process significantly improves \PaperAcronym{}'s ability to manage complex tasks and generate accurate G-code for intricate designs.

\section{Self-Corrective G-Code Generation}\label{sec:self_correction}
%
In this section, we introduce the self-corrective G-code generation mechanism. In general, the self-corrective code generation approach moves beyond the limitations of a simple prompt-and-response paradigm, where the model generates code in a single step without further refinement. Instead, we adopt a more dynamic and iterative flow paradigm. In this approach, the model's initial code generation is treated as a starting point. This initial code is then subjected to a series of automated tests designed to identify potential errors or deviations from the user's intent. The results of these tests are then fed back into the model, allowing it to \quotes{reflect} on its output and make necessary corrections. This iterative process of code generation, testing, and refinement continues until a valid and accurate code is produced, or a predetermined maximum number of iterations is reached. This ensures both a high level of accuracy and a bounded execution time, preventing infinite loops in cases where a perfect solution might be unattainable.

\begin{figure}
    \centering
    \includegraphics[width=1\linewidth]{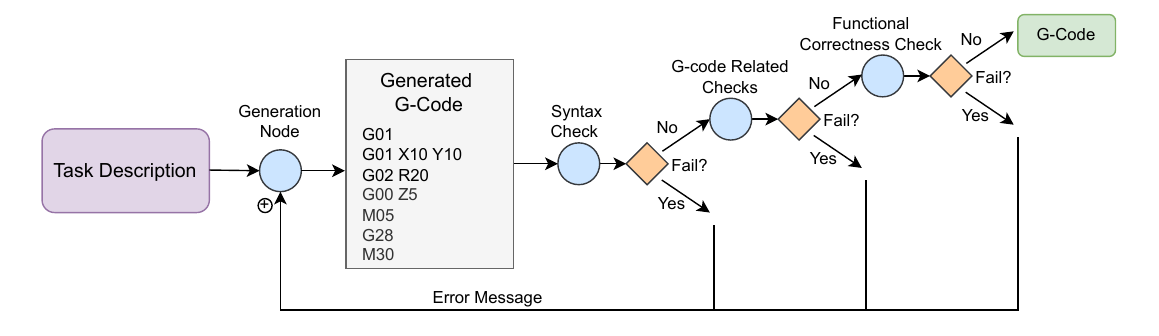}
    \caption{Self-corrective G-Code generation graph}
    \label{fig:graph}
\end{figure}

Figure~\ref{fig:graph} illustrates the application of the self-corrective code generation approach to the specific task of G-code generation. The process begins with a user-provided task description, which serves as input to the Generation Node. This node, representing our trained LLM model, produces an initial draft of the G-code. However, instead of assuming this initial output is correct, the system embarks on a series of validation checks. The generated G-code is first subjected to a syntax check, ensuring adherence to the grammatical rules of the G-code language. If errors are detected, an error message is generated, and the model uses this feedback to refine the code. If the code passes the syntax check, it progresses to a series of G-code-related checks, which assess the code's logical consistency and adherence to machining conventions. Finally, a functional correctness check evaluates whether the generated G-code, if executed, would produce the desired outcome as specified in the user's task description. If any check fails, the error message is routed back to the Generation Node, prompting the model to revise the code. In the following, we delve deeper into the specific syntactic and semantic tests employed within this framework.

\subsection{G-code Validation}
Syntactic validation in our self-corrective code generation framework focuses on ensuring that the generated G-code adheres to the strict grammatical rules of the G-code language. This process involves a three-step approach. First, the generated G-code program is parsed line by line. Then, within each line, individual G-code commands are extracted. These commands typically consist of a letter (e.g., G, M) followed by a numerical code (e.g., G01, M03). Finally, each extracted command is compared against a comprehensive list of recognized G-code instructions. Any command that cannot be matched to this list is flagged as syntactically incorrect. For example, a command like \quotes{G022}, intending to perform a circular interpolation, would be deemed syntactically incorrect due to the extraneous \quotes{2} at the end, which deviates from the standard \quotes{G02} format for circular interpolation.

Aside from syntactic errors, detecting unreachable code is crucial for ensuring the generated G-code program executes as intended. \PaperAcronym{} incorporates a dedicated check to identify and flag such logical errors in the program flow. Specifically, the system analyzes the sequence of G-code commands, identifying the program's end by locating the \quotes{M30} command, which signifies program termination. Any G-code commands appearing after the \quotes{M30} command are inherently unreachable, as the program would have already terminated. These lines are flagged as potential errors, prompting the model to revise the code and ensure all commands are logically reachable within the program's flow.

Our G-code validation process includes a crucial safety check to prevent potentially dangerous rapid movements while the cutting tool is engaged. This check operates by continuously monitoring the tool's state within the G-code program. First, the system determines if the tool is currently engaged in a cutting operation by tracking the activation of relevant commands, such as those initiating spindle rotation and feed. Next, the system identifies any rapid movement commands, specifically \quotes{G0} or \quotes{G00,} which instruct the machine to move the tool at its maximum speed. If a rapid movement command is detected while the tool is actively cutting, the system flags a potential safety violation. This proactive measure ensures the generated G-code prioritizes safety by preventing abrupt, high-speed movements that could damage the workpiece, the cutting tool, or even the machine itself.

To ensure the safety and accuracy of drilling operations, our G-code validation framework incorporates a specialized \quotes{Validate Safe Drilling} check. This check analyzes the generated G-code to guarantee the machine drills only at the specified depths and avoids any unintended horizontal movements at those depths, which could lead to tool breakage or damage to the workpiece. The validation process involves four key steps: First, the G-code program is analyzed line by line. Second, the system continuously tracks the tool's current position in all three axes (X, Y, and Z). Third, any horizontal movements, indicated by changes in the X or Y coordinates, are identified. Finally, the system flags any invalid horizontal movements that occur below a predefined safe height, ensuring the tool only moves horizontally when it is safely above the workpiece or at a designated clearance level. This rigorous validation process ensures that drilling operations are executed safely and precisely, preventing potential errors that could compromise the integrity of the workpiece or the drilling tool.

\subsection{Functional Correctness}
Functional correctness validation of the generated G-code is paramount for ensuring the output aligns with the user's intended machining outcome. This validation process centers on a comparative analysis between the toolpath generated from the G-code program and the user's explicitly defined specifications. Algorithm~\ref{alg:functional} outlines the procedure for validating the functional correctness of the generated G-code. The algorithm begins by extracting coordinate data from both the generated G-code (line~\ref{l:1}) and the user-provided parameters (lines~\ref{l:2}-\ref{l:3}). If the user has specified a starting point and toolpath, these are extracted and, if necessary, combined to form a complete user-defined toolpath. Both the G-code-derived toolpath and the user-defined toolpath are then refined by removing duplicate points (lines~\ref{l:4}-\ref{l:5}).
	\begin{algorithm}
		\caption{Validate Functional Correctness of G-Code}
        \label{alg:functional}
		\begin{algorithmic}[1]
			\Require generated G-Code $gcode\_string$, $parameters\_string$, tolerance
			\Ensure distance $d$ between user-verified and generated paths
			\State Extract coordinates $X_g$, $Y_g \gets$ Parse($gcode\_string$)
			\State Construct G-code tool path $gcode\_path \gets$ ConstructPath($X_g$, $Y_g$) \label{l:1}
			\State Extract $user\_params$ $\gets$ ParseExtractedParameters($parameters\_string$) \label{l:2}
			\If{$user\_params$ exist} 
			\State Get user-defined starting point $st \gets$ $user\_params[\text{starting\_point}]$ 
			\State Get user-defined tool path $user\_path$ $\gets$ $user\_params[\text{tool\_path}]$ 
			\State Get $X_u$, $Y_u \gets$ ExtractCoordinates($user\_path$)
			
			\If{$user\_path$ exists AND $st$ $ \ne user\_path[0]$}
			\State Prepend $st$ to user-defined $X_u$ coordinates
			\State Prepend $st$ to user-defined $Y_u$ coordinates 
			\EndIf
			
			\State Construct user-defined tool path $user\_path \gets$ ConstructPath($X_u$, $Y_u$) \label{l:3}
			\State Estimate $gcode\_path \gets$ RemoveDuplicates($gcode\_path$) \label{l:4}
			\State Estimate $user\_path \gets$ RemoveDuplicates($user\_path$)	\label{l:5}		
			\State Calculate $d \gets$ HausdorffDistance($gcode\_path$, $user\_path$) \label{l:6}	
			
			\If{calculated distance $d \leq$ tolerance}
			\State Message $\gets$ "tool paths match within tolerance"
			\Else
			\State Message $\gets$ "tool paths do not match" \label{l:7}	
			\EndIf
			\EndIf 
			\State \Return Message, Distance $d$
		\end{algorithmic}
	\end{algorithm}

The core of the functional correctness validation lies in quantifying the geometric similarity between the user-verified toolpath and the toolpath generated from the LLM's output (lines~\ref{l:6}-\ref{l:7}). For this purpose, we leverage the Hausdorff distance, a well-established metric in computer vision and shape analysis for comparing sets of points. The Hausdorff distance represents the maximum distance between a point in one set to its nearest neighbor in the other set, providing a robust measure of dissimilarity between two geometric shapes. A predefined threshold for the Hausdorff distance is established based on the desired level of accuracy for the specific machining task. If the calculated Hausdorff distance between the two toolpaths falls below this threshold, the generated G-code is deemed functionally correct, indicating a high degree of similarity. Conversely, if the distance exceeds the threshold, an error signal, incorporating the calculated Hausdorff distance value, is appended to the original input prompt and fed back to the LLM. This error-augmented feedback loop iteratively guides the LLM to refine its output, generating a G-code program that produces a toolpath more closely aligned with the user's specifications. This iterative process continues until the generated G-code successfully produces a toolpath within the acceptable Hausdorff distance threshold or a maximum number of iterations is reached.


\section{Performance Evaluation}\label{sec:evaluation}

the main intention is to enhance open-source models to achieve comparable performance to proprietary models. to this end, we leverage structured prompts and self-corrective mechanisms.

\subsection{Experimental Setup}

In this section, we introduce our experimental environment. We leveraged open-source language models hosted on the Hugging Face Hub for our implementation. Specifically, We evaluate \PaperAcronym{} using four distinct LLMs: the proprietary GPT-3.5 and three open-source models, Zephyr, Starcoder, and CodeLlama. Zephyr and CodeLlama, two 7B parameter models, have been employed in their pre-trained form. For Starcoder, we opted for the 3B parameter variant and further fine-tuned it to enhance its performance on CNC-related tasks. The fine-tuning process leveraged the Stack dataset, a publicly available collection of code from GitHub repositories, including G-code, Python, and various other programming languages. To accelerate training, we implemented PEFT and mixed precision training techniques. The fine-tuned model was subsequently uploaded to Hugging Face, enabling efficient inference through their model serving infrastructure.

We implemented the self-corrective mechanism and RAG using LangChain's LangGraph framework. LangGraph allowed us to create a directed acyclic graph (DAG) of language model calls, enabling iterative refinement of outputs and integration of external knowledge. Our RAG mechanism begins by loading and processing text from the provided PDF files that contain CNC domain knowledge. Then, it generates embeddings for these text elements using OpenAI's embedding model. These embeddings are stored in a FAISS vector store, which allows for efficient similarity searches. A retrieval-augmented QA chat prompt is pulled from a specified hub and used to create a document combination chain. Finally, a retrieval QA chain is constructed by linking the retriever and document combination chain, and this chain is returned for use in question-answering tasks.

We developed a user-friendly web interface using Streamlit, a Python library for creating interactive web applications. The interface allows users to: (1) Input task descriptions, (2) Validate extracted parameters, (3) Simulate user-defined requirements, and (4) Generate and display G-code. The web application communicates with our fine-tuned model and LangGraph pipeline via API calls, ensuring real-time responsiveness. For G-code simulation, we leverage CAMotics, an open-source CNC simulator. While the G-code is generated within our web interface, users must download the code and manually import it into the CAMotics environment for visualization. This step allows for a comprehensive 3D simulation of the CNC machining process. To provide an immediate visual representation of the executed task, we implemented a 2D visualization feature directly within the web interface. This functionality is powered by a custom Python script that interprets the generated G-code and renders a 2D projection of the tool path.

\begin{figure*}[htbp]
	\centering
	\subfloat[Task \#1]{\label{fig:square}\includegraphics[width=0.15\linewidth]{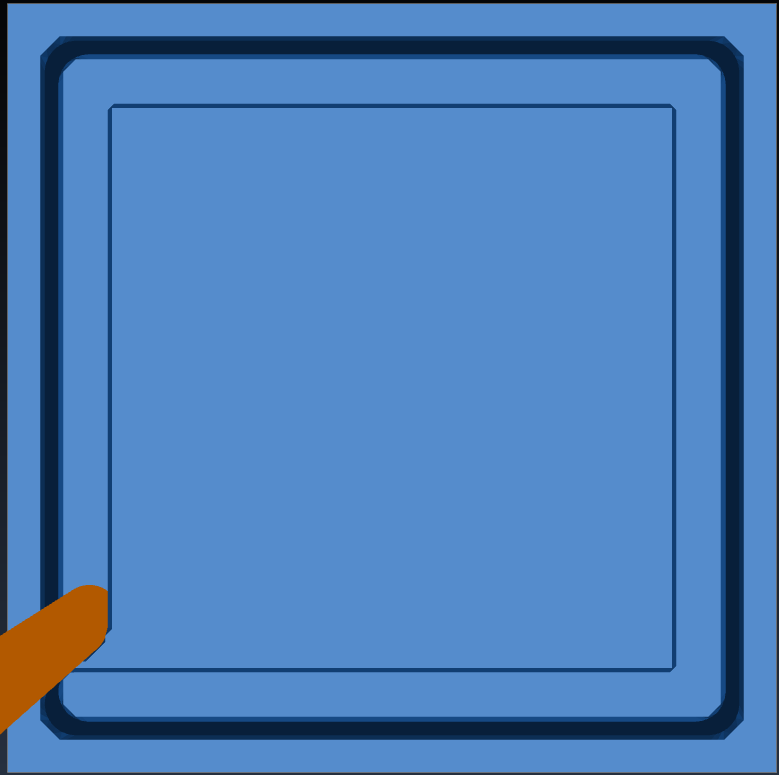}} \hspace{1mm}
	\subfloat[Task \#2]{\label{fig:hexagon}\includegraphics[width=0.15\linewidth]{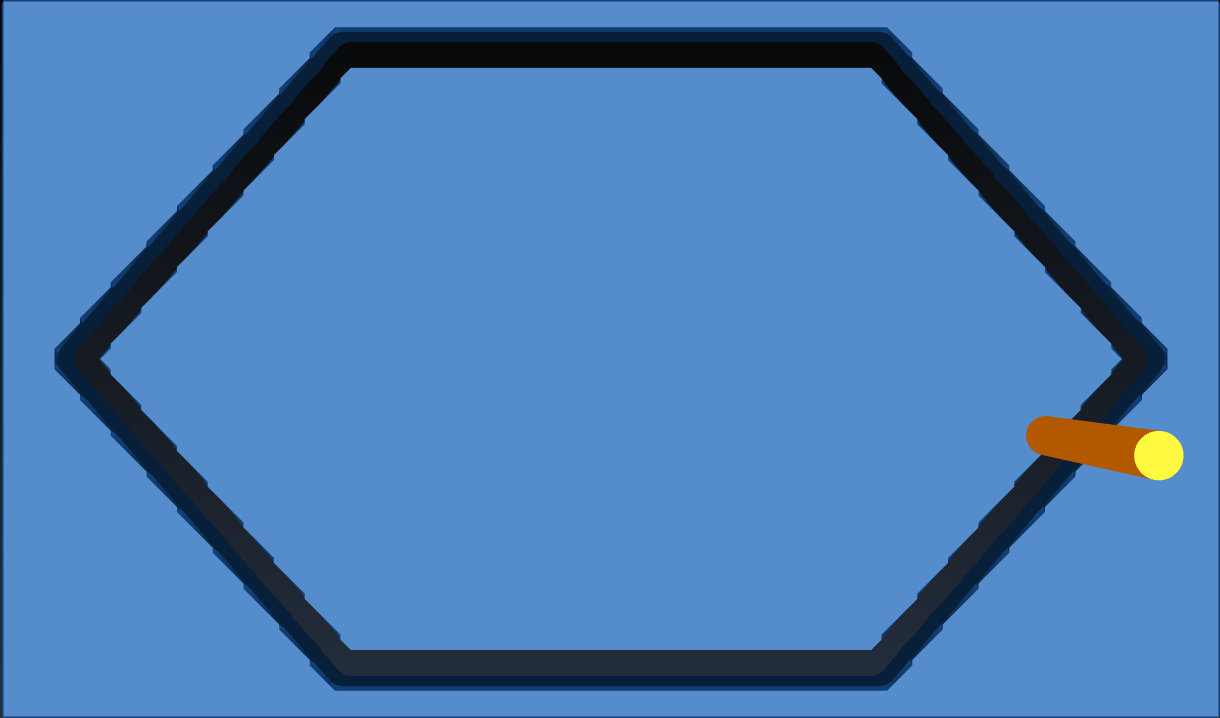}} \hspace{1mm}
 	\subfloat[Task \#3]{\label{fig:custom}\includegraphics[width=0.15\linewidth]{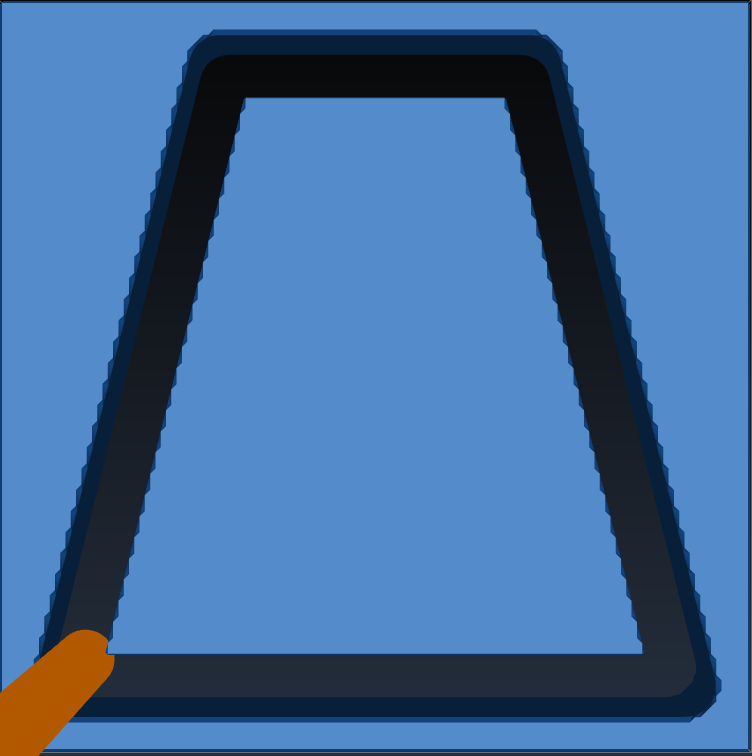}} \hspace{1mm}
    \subfloat[Task \#4]{\label{fig:circle}\includegraphics[width=0.15\linewidth]{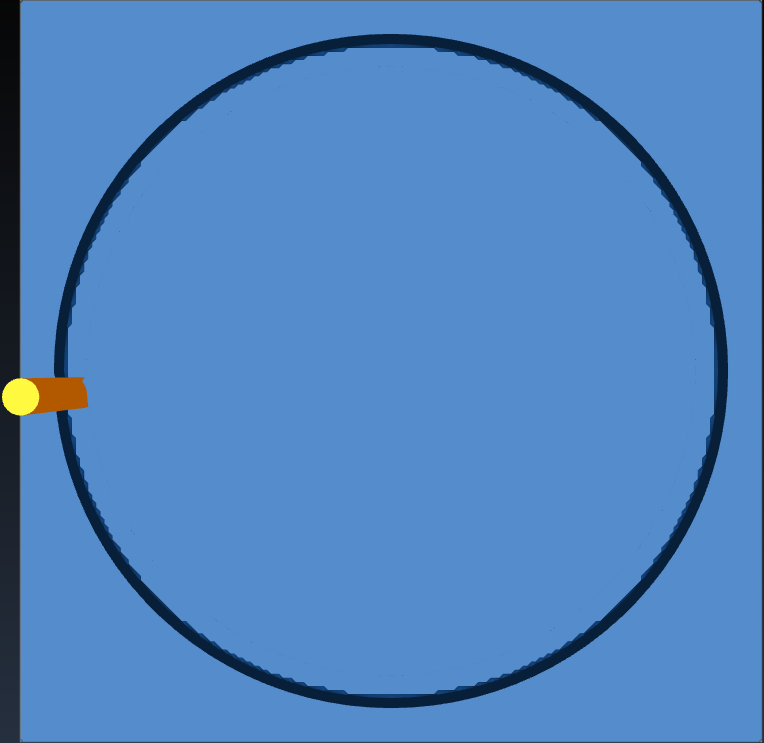}} \hspace{1mm}
    \subfloat[Task \#5]{\label{fig:pocket}\includegraphics[width=0.15\linewidth]{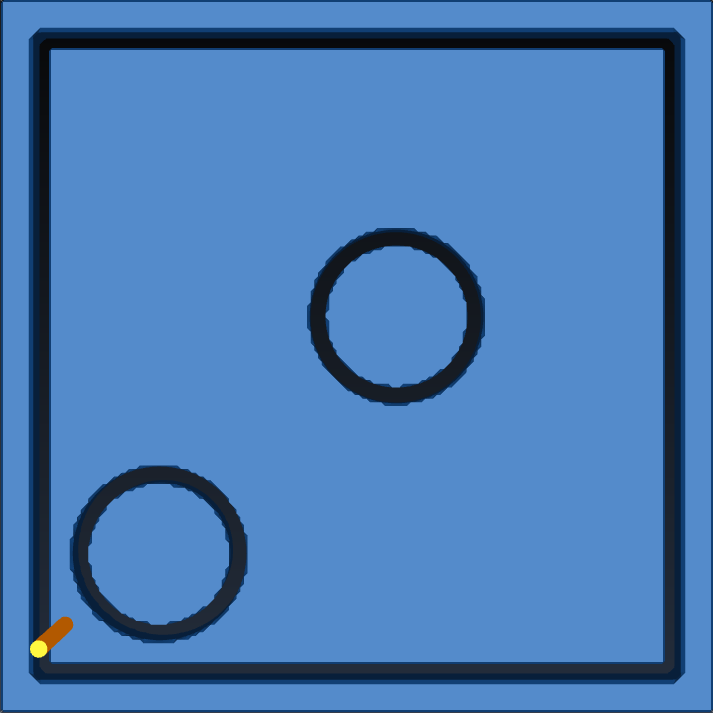}} \hspace{1mm}
 	\subfloat[Task \#6]{\label{fig:matrix}\includegraphics[width=0.15\linewidth]{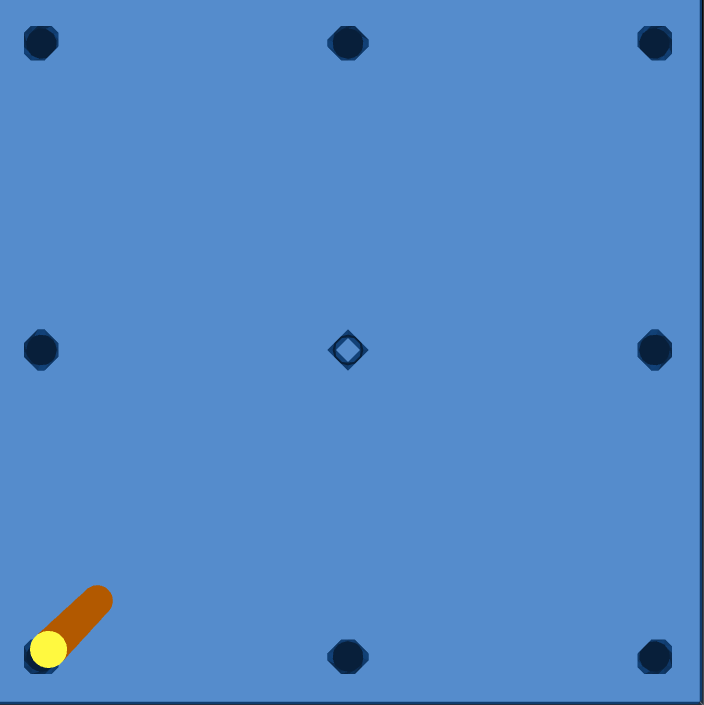}} \hspace{1mm}
    \caption{Simulation of generated G-codes for the various test cases}
	\label{fig:simulations} 
\end{figure*}

In our experiments, we investigated the efficacy of LLMs for G-code generation, focusing on the impact of RAG, structured prompts, and our novel self-corrective mechanism. Specifically, our evaluation addresses three key research questions: (1) Does RAG exclusively benefit LLM performance when using unstructured prompts? (2) How does the use of parameter extraction for structured prompt creation affect performance across various LLMs and tasks? (3) How does the self-corrective mechanism's performance vary across tasks of differing complexity? To answer these questions, We evaluated performance across six distinct G-code generation tasks, each targeting the creation of a different geometric shape. These tasks, illustrated in Figure~\ref{fig:simulations}, range in complexity and include milling a square, hexagon, irregular shape, circle, and a rectangular pocket with two circular islands, as well as drilling a grid of holes.

LLM performance was quantified using two primary metrics: success rate and average iterations. The success rate is defined as the proportion of trials where the generated G-code passed all validation tests. These tests ensure the generated code accurately represents the target geometry and adheres to machining constraints. Average iterations, representing the number of correction cycles invoked by the self-corrective mechanism before producing a valid G-code. This metric provides insight into both model responsiveness to error feedback and the inherent complexity of each task. All reported results represent the average of five independent experimental runs.

\subsection{Experimental Results}

\begin{figure*}[htbp]
	\centering
	\subfloat[Impact of RAG]{\label{fig:rag}\includegraphics[width=0.33\linewidth]{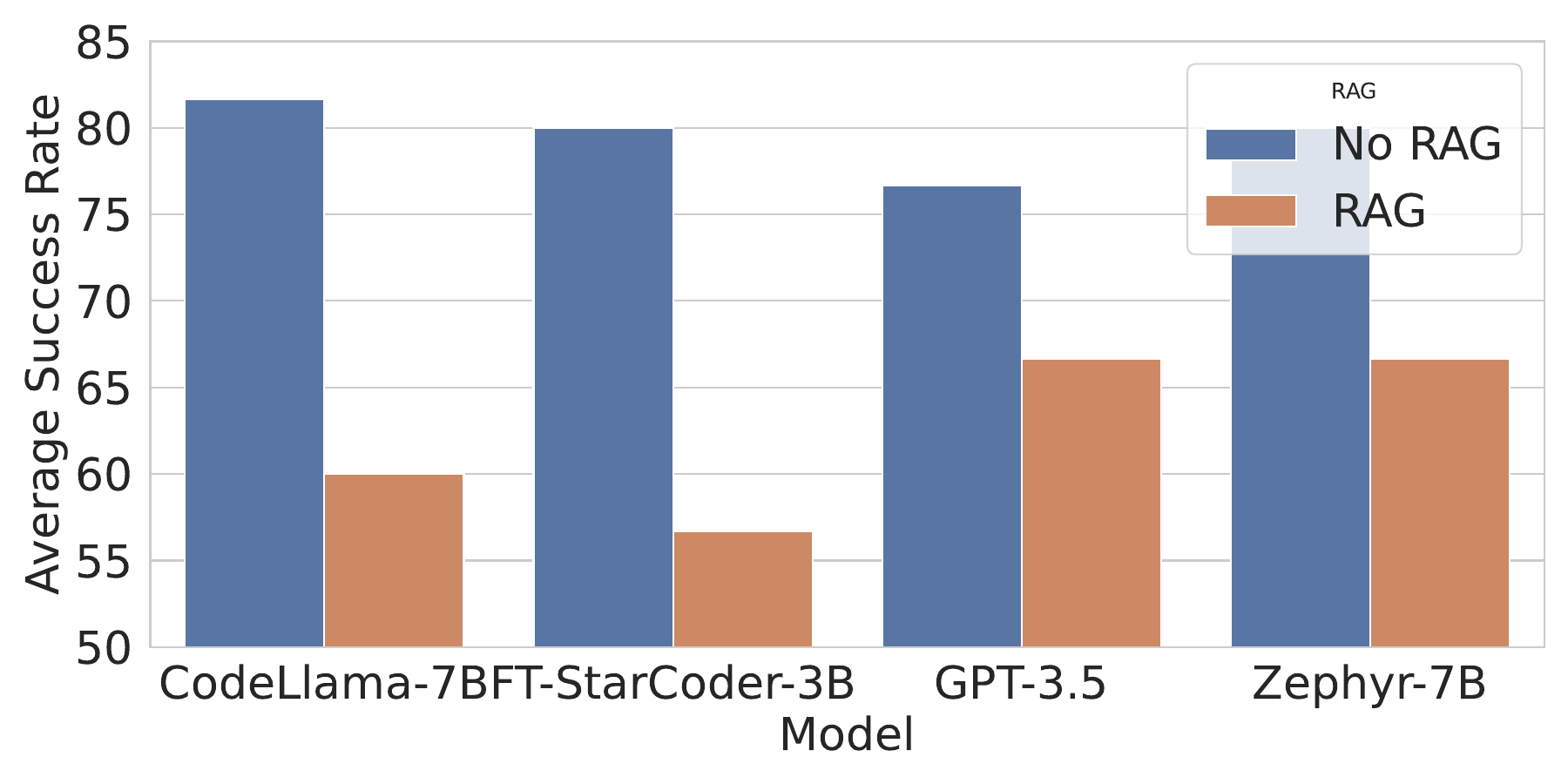}} 
	\subfloat[Structured prompts]{\label{fig:heatmap}\includegraphics[width=0.33\linewidth]{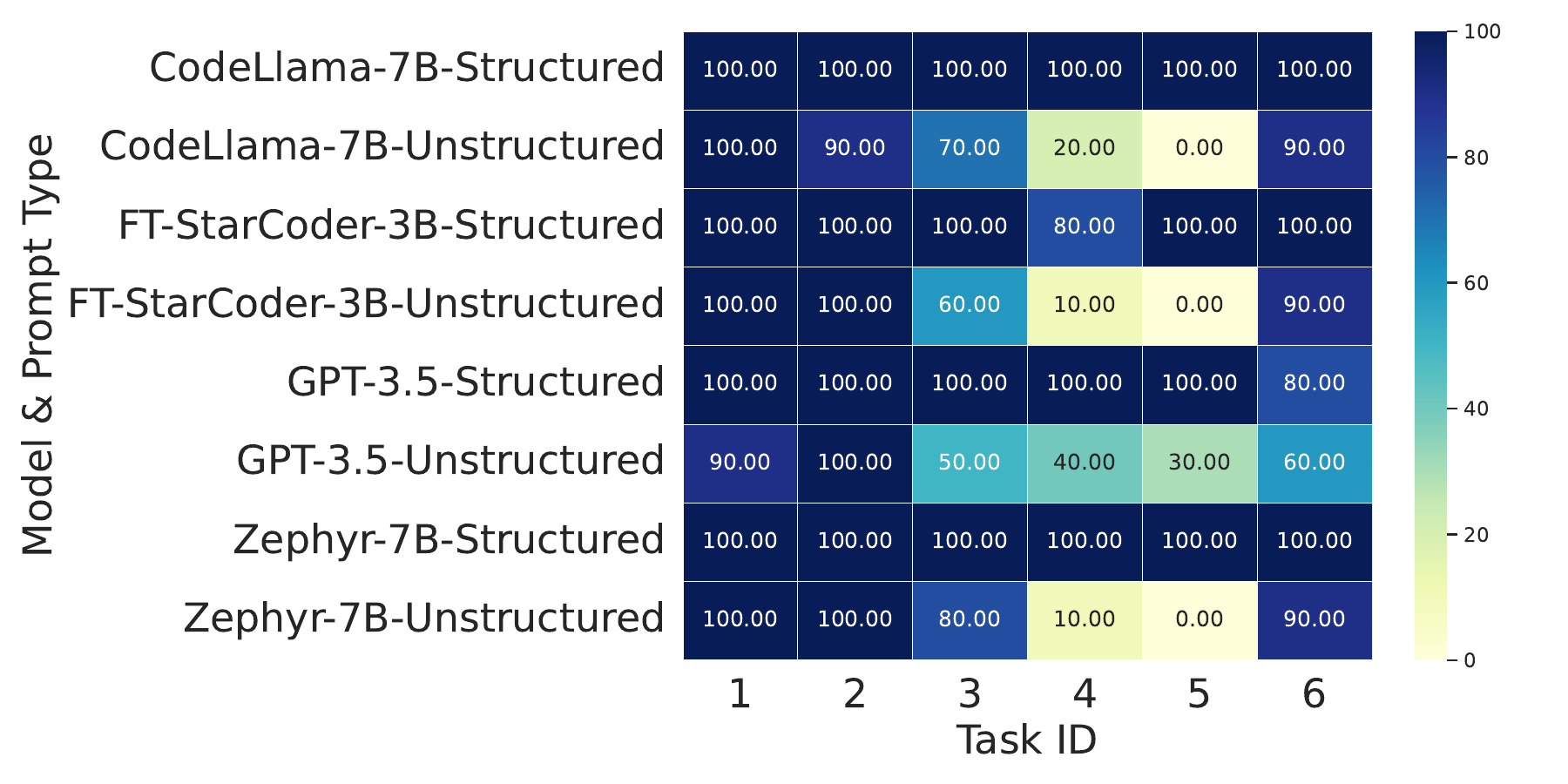}} 
 	\subfloat[Task difficulty]{\label{fig:iterations}\includegraphics[width=0.33\linewidth]{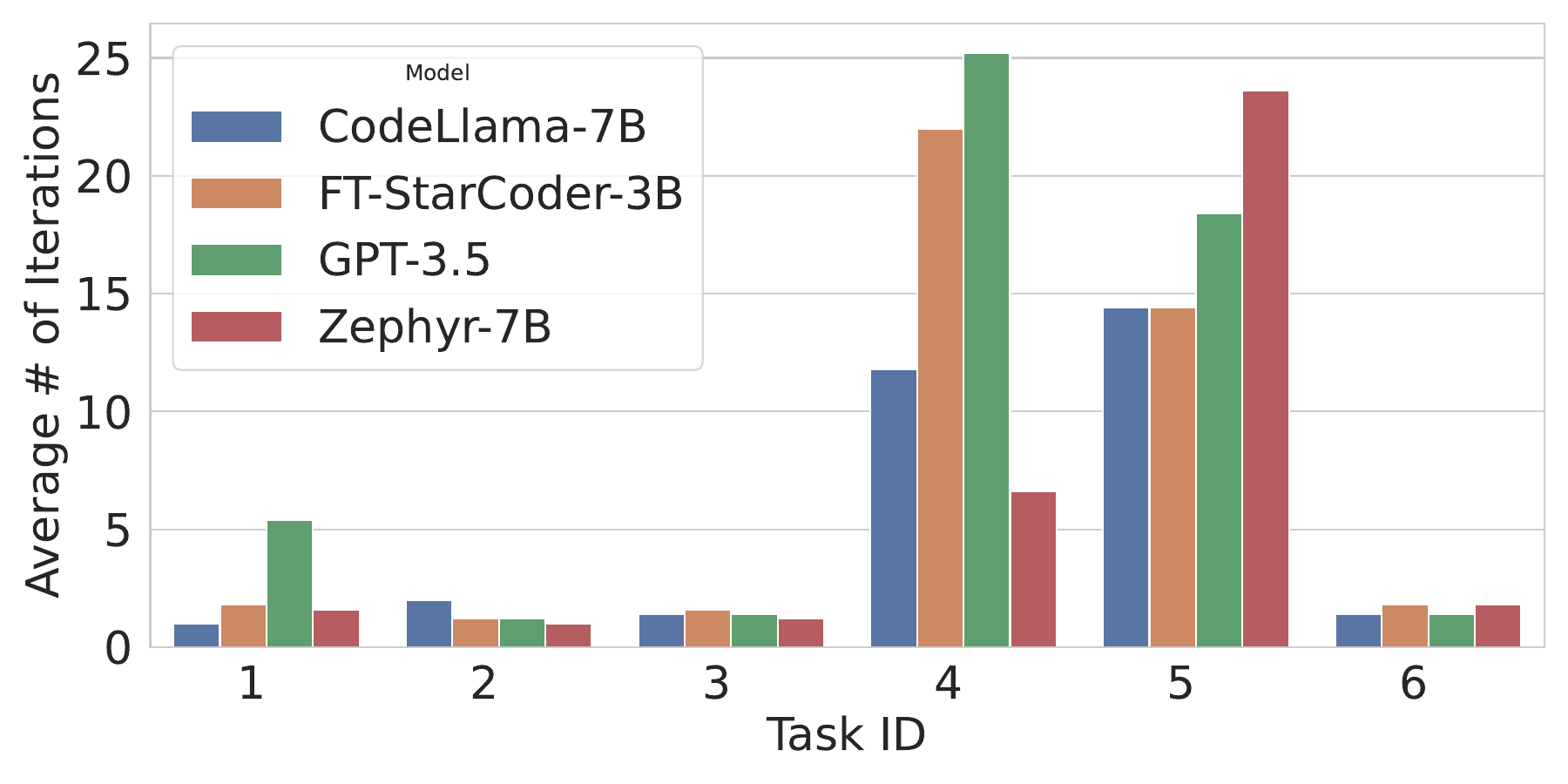}} 
    \caption{Evaluation results of \PaperAcronym{}}
	\label{fig:results} 
\end{figure*}

In this section, we introduce the results of our experiments. For instance, Figure~\ref{fig:rag} illustrates the average success rate of several LLM models when generating G-code using unstructured prompts, both with and without the assistance of RAG. The models evaluated include CodeLlama-7B, FT-StarCoder-3B, GPT-3.5, and Zephyr-7B. The results reveal that all models achieve higher success rates without RAG, with CodeLlama-7B and FT-StarCoder-3B reaching approximately 80\% success. In contrast, the incorporation of RAG results in a noticeable decrease in performance across all models, suggesting that RAG may not be beneficial for improving success rates with unstructured prompts in this context. This indicates that the models are more effective in handling unstructured prompts independently, and RAG may introduce complexities that hinder their performance.

Figure~\ref{fig:heatmap} shows a heatmap that illustrates the performance of various language LLMs using both structured and unstructured prompts across six tasks. Each cell shows the success rate, representing the percentage of times the model successfully generated an accurate G-code for the given task. Darker shades correspond to higher success rates, with structured prompts generally yielding better results across all models. Notably, structured prompts consistently achieve a 100\% success rate for most tasks, indicating their effectiveness in guiding models to generate correct outputs. In contrast, unstructured prompts result in more variability, with success rates dropping significantly for tasks 3 and 4, particularly for CodeLlama-7B and GPT-3.5. This suggests that while some models can handle unstructured prompts effectively, others benefit significantly from the clarity provided by structured guidance.

Finally, the bar chart in Figure~\ref{fig:iterations} displays the average number of iterations required by the four LLMs across six tasks. Tasks 4 and 5 appear to be the most challenging, with all models requiring significantly more iterations, especially GPT-3.5, which has the highest count. For task 5, its complexity emerges from the need to mill three different shapes. In contrast, tasks 2, 3, and 6 show minimal iteration requirements, indicating these tasks are less complex for the models. Interestingly, Zephyr-7B and FT-StarCoder-3B exhibit similar performance across most tasks, with slight variations in iteration counts. CodeLlama-7B consistently requires fewer iterations than the other models for task 1, suggesting greater efficiency in simpler tasks. The results highlight how task complexity can influence iteration needs, with notable differences in how each model handles varying task demands.

The evaluation of \PaperAcronym{} in G-code generation reveals several key insights. Structured prompts consistently outperform unstructured ones across all models and tasks, with CodeLlama-7B and FT-StarCoder-3B achieving perfect success rates when using structured inputs. This underscores the importance of well-formulated prompts in enhancing model accuracy. Interestingly, the incorporation of RAG with unstructured prompts decreased performance. This unexpected result warrants further investigation into the optimal integration of RAG for G-code generation tasks. The complexity of tasks significantly influences model performance, as evidenced by the varying number of iterations required across different tasks. Notably, tasks 4 and 5 emerged as the most challenging, requiring substantially more iterations from all models. This variability in task difficulty highlights the ability of \PaperAcronym{} to handle diverse G-code generation scenarios. Overall, the results highlight the ability of structured prompts and the self-corrective G-code generation mechanism for G-code generation tasks.


\section{Conclusion}\label{sec:conclusion}

In this paper, we introduced \PaperAcronym{}, a system capable of interpreting natural language instructions and producing executable G-code. While the paper focuses on comparing open-source models to proprietary ones, the results suggest that with structured prompts and self-corrective mechanisms, open-source models can achieve comparable performance. These components ensure the reliability and safety of the generated code, addressing critical concerns in manufacturing environments. As \PaperAcronym{} continues to evolve, it has the potential to significantly reduce the time and expertise required for CNC programming, ultimately enhancing efficiency and innovation in the manufacturing sector. Future research should explore optimizing RAG integration and further refining prompt structures to maximize the potential of LLMs in complex machining applications.

\bibliography{main}

\begin{thebibliography}{{Au}14}

\bibitem[{Au}14]{autodesk2014fundamentals}
{Autodesk, Inc.}: Fundamentals of CNC Machining: A Practical Guide for Beginners.
\newblock Autodesk, Inc., 2014.
\newblock Desk Copy, Document Number: 060711.

\bibitem[Au24]{fusion2024}
Autodesk Fusion: More than CAD, it's the future of design and manufacturing.
\newblock \url{https://www.autodesk.com/products/fusion-360/}, Accessed: 01 July 2024.

\bibitem[Cr24]{mastercam2024}
Creating software and services that solve the world’s manufacturing challenges.
\newblock \url{https://www.mastercam.com/}, Accessed: 15 July 2024.

\bibitem[Si08]{siemens2008shopmill}
Siemens: .
\newblock SINUMERIK 840D sl ShopMill Operation/Programming.
\newblock Siemens AG, 2008.
\newblock © Siemens AG 2008.

\bibitem[Ul24]{cura2024}
UltiMaker Cura.
\newblock \url{https://ultimaker.com/software/ultimaker-cura/}, Accessed: 01 July 2024.

\end{thebibliography}
\end{document}